\documentclass{jfm}

\usepackage{graphicx}
\usepackage{bm}
\usepackage{newtxtext}
\usepackage{newtxmath}
\usepackage{natbib}
\usepackage{amsmath}
\usepackage{empheq}
\usepackage{hyperref}
\usepackage{subfigure}
\usepackage{dcolumn}
\usepackage{physics}
\usepackage{tcolorbox}

\RequirePackage{color}
\hypersetup{
  colorlinks = true,
  urlcolor   = blue,
  citecolor  = black,
}

\newcommand{\RomanNumeralCaps}[1]
\linenumbers

\title{Inertial range scaling of inhomogeneous turbulence}

\author{
  Ryo Araki\aff{1,2}
  \and Wouter J. T. Bos\aff{1}
  \corresp{\email{wouter.bos@ec-lyon.fr}}
}

\affiliation{
  \aff{1} Univ Lyon, CNRS, Ecole Centrale de Lyon, INSA Lyon, Univ Claude Bernard Lyon 1,
  LMFA,
  UMR5509, 69340 Ecully, France
  \aff{2} Graduate School of Engineering Science,
  Osaka University,
  1-3 Machikaneyama, Toyonaka, Osaka 560-8531, Japan
}

\begin{document}
\maketitle

\begin{abstract}
{\bf Abstract:} We investigate how inhomogeneity influences the \(k^{-5/3}\) inertial range scaling of turbulent kinetic energy spectra (with $k$ the wavenumber).
For weak statistical inhomogeneity, the energy spectrum can be described as an equilibrium spectrum plus a perturbation.
Theoretical arguments suggest that this latter contribution scales as $k^{-7/3}$.
This prediction is assessed using direct numerical simulations of three-dimensional Kolmogorov flow.
\end{abstract}

%

\section{Introduction}

Kolmogorov postulated in 1941 that the small scales of turbulent flows away from boundaries can be considered universal if the Reynolds number is sufficiently large~\citep{Kolmogorov1941_the_local}.
The small scales are then supposed to be in equilibrium, and the energy spectrum satisfies,
\begin{equation}\label{eq:E}
  E(k,\vb*{x}, t) \sim \epsilon(\vb*{x}, t)^{2/3} k^{-5/3},
\end{equation}
where $\epsilon$ is the average energy dissipation rate.
For this expression to hold, the wavenumber $k$ should be sufficiently large compared to $L(\vb*{x}, t)^{-1}$, the inverse of the length scale characterising the largest scales of the flow, and sufficiently small compared to the inverse of the Kolmogorov-scale $\eta(\vb*{x}, t)^{-1}$ (with $\eta=\nu^{3/4} \epsilon^{-1/4}$), associated with the smallest scale of the flow.

In expression~\eqref{eq:E}, the time and space dependence of $E$ and $\epsilon$ need some particular attention.
Theoretically, the most convenient flow type for investigating inertial range scaling is an infinitely large and statistically stationary flow without boundaries.
Since all practical flows are limited in size and lifetime, the dissipation rate will be dependent, even on average, on either position $\vb*{x}$ or time $t$, or both.
Expression~\eqref{eq:E} will therefore hold only locally in subdomains of space and time-intervals large enough compared to the considered length and time scales.

Indeed, the assumptions allowing the simple prediction~\eqref{eq:E} are that the scales $k$ can be considered locally isotropic, stationary, and homogeneous.
The criterion $k \gg L^{-1}$ represents the implicit assumption that the influence of anisotropy, instationarity and inhomogeneity decreases as a function of scale.
The present investigation aims to substantiate this assumption, particularly regarding the influence of inhomogeneity.

As an illustration, let us discuss the influence of statistical instationarity on the behaviour of the small scales.
This subject was addressed by~\citet{Yoshizawa1994_nonequilibrium}, who proposed that the influence of instationarity at large wavenumbers can be described as a perturbation on the energy spectrum as
\begin{equation}\label{eq:E01}
  E(k,\vb*{x}, t) = E_0(k,\vb*{x}, t) + E_1(k,\vb*{x}, t),
\end{equation}
where the equilibrium part $E_0$ is given by~\eqref{eq:E} and the perturbation scales as
\begin{equation}\label{eq:E1_temporal}
  E_1^{T}(k,\vb*{x}, t) = C_Y \dv{\epsilon(\vb*{x}, t)}{t} \epsilon(\vb*{x}, t)^{-2/3} k^{-7/3},
\end{equation}
where the superscript $T$ denotes that we consider perturbations due to instationarity.
Numerical evidence of this scaling was first obtained by~\citet{Horiuti2011_multimode} for the case of homogeneous shear flow and by~\citet{Horiuti2013_nonequilibrium} for statistically isotropic turbulence in a periodic box.
Further theoretical discussion and a more straightforward derivation of~\eqref{eq:E1_temporal} can be found in~\citet{Rubinstein2005_self-similar, Woodruff2006_multiple-scale, Bos2017_dissipation}.

A similar approach is applied in various other configurations where the influence of external effects on isotropic turbulence is modelled as a perturbation to the energy spectrum.
The effect of a mean-shear on isotropic turbulence was treated perturbatively by~\citet{Ishihara2002_anisotropic}.
Stratified turbulence was considered in~\citet{Kaneda2004_small-scale} and the limit of weakly compressible turbulence by~\citet{Bertoglio2001_two-point}.


The effect of large-scale temporal fluctuations on the kinetic energy spectrum
is thus proportional to $k^{-7/3}$, which decays more rapidly than the equilibrium spectrum~\eqref{eq:E} with the \(k^{-5/3}\) scaling.
In the remainder of this investigation, we will focus on the influence of inhomogeneity on the scaling of turbulent kinetic energy, which has received little attention.
Using Karhunen-Loeve eigenfunctions, it was illustrated that Kolmogorov's equilibrium spectrum can be observed in statistically inhomogeneous flows~\citep{Knight1990_Kolmogorov,Moser1994_Kolmogorov,Liao2015_Kolmogorov}.
By using the SO(3) symmetry group decomposition, \citet{Kurien2000_scaling} showed that structure functions contain a subdominant scaling component associated with inhomogeneity.

We further assess at which rate statistical homogeneity is recovered at small scales.
To that, in \S~\ref{sec:Derivation} we derive an analytical prediction of the scaling of \(E_1^X(k,\vb*{x})\), where the superscript \(X\) denotes the perturbation due to inhomogeneity, in stationary inhomogeneous turbulence.
This expression will be the inhomogeneous equivalent of equation~\eqref{eq:E1_temporal}.
In \S~\ref{sec:Assessment}, we report the results of Direct Numerical Simulations (DNS) of the three-dimensional Kolmogorov flow to assess the predictions.
Section~\ref{sec:Conclusion} concludes this investigation.

\section{Derivation of the spectral correction due to inhomogeneity \label{sec:Derivation}}

The main difficulty in the present investigation comes from the fact that we investigate a multi-scale description (the energy spectrum) in an inhomogeneous setting.
To simplify as far as possible, we restrict ourselves to a fairly simple setting, where the (statistical) inhomogeneity is periodic in space, and the flow is stationary and far away from boundaries.
Before addressing the inhomogeneous multi-scale description, we will first consider the pointwise energy balance of the flow.

\subsection{Kinetic-energy budget in inhomogeneous turbulence \label{subsec:Kinetic-energy}}

We consider a statistically inhomogeneous flow kept in a statistically stationary state by a steady forcing $f(z)$.
The forcing in the present manuscript consists of a unidirectional steady body force in the $x$-direction with a sinusoidal dependence in the $z$-direction.
The Navier-Stokes equations for this specific system write
\begin{equation}\label{eq:NSE}
  \frac{\mathrm{D} \vb*{\mathcal{U}} (\vb*{x}, t)}{\mathrm{D} t}
    = -\grad{\mathcal{P}} (\vb*{x}, t)
    + \nu \Delta \vb*{\mathcal{U}} (\vb*{x}, t)
    + f(z) \vb*{e}_x,
\end{equation}
where \(\mathrm{D} / \mathrm{D}t\) is the material derivative, $\mathcal{P}$ is the pressure (divided by density) ensuring incompressibility $\div{\vb*{\mathcal{U}}} = 0$, and \(\vb*{e}_x\) denotes the unit vector in the \(x\)-direction.

The equations for the mean flow and the kinetic energy of the fluctuations can be derived by introducing the Reynolds decomposition \(\vb*{\mathcal{U}} = \expval{\vb*{\mathcal{U}}} + \vb*{u}\), where \(\expval{\vb*{\mathcal{U}}}\) is the ensemble-averaged velocity and \(\vb*{u} = (u, v, w)\) the fluctuation.
The specific forcing considered in the present investigation leads to a mean flow $\expval{\vb*{\mathcal{U}} (\vb*{x}, t)} = U(z) \vb*{e}_x$.
Then, the kinetic energy corresponding to the mean flow can be written as
\begin{equation}
  K_U(z) = U(z)^2/2,
  \label{eq:K_Uz}
\end{equation}
and the kinetic energy of the fluctuations is
\begin{equation}\label{eq:K}
  K(z) = \frac12 \qty[\expval{u^2} (z) + \expval{w^2} (z) + \expval{w^2} (z)].
\end{equation}
The equation for the mean-velocity $U(z)$ reduces to,
\begin{equation}
  \frac{\mathrm{D} U(z)}{\mathrm{D} t}
    = -\pdv{z} \expval{u w} (z)
    + f(z)
    + \nu \pdv[2]{U(z)}{z}
    = 0.
\end{equation}
The details are, for instance, provided in~\cite{Bos2020_production}.
The equation for the turbulent kinetic energy writes, in a steady state,
\begin{equation}\label{eq:dKdt}
  \frac{\mathrm{D} K(z)}{\mathrm{D} t}
    = p(z) - \epsilon(z) + d(z)
    = 0,
\end{equation}
where the production $p(z)$, dissipation $\epsilon(z)$, and diffusion $d(z)$ terms are given by
\begin{empheq}{align}
  p(z)
    &= -\expval{u w} (z) \pdv{U(z)}{z},
  \label{eq:pz} \\
  \epsilon(z)
    &= \nu \expval{\pdv{u_i}{x_j} \pdv{u_i}{x_j}} (z),
  \label{eq:epsilonz} \\
  d(z)
    &= -\pdv{z} \qty(\expval{\mathcal{P} w} (z)
    + \expval{u_i u_i w} (z)
    - \nu \pdv{K(z)}{z}),
  \label{eq:dz}
\end{empheq}
respectively.
The first term $p(z)$ represents the production of turbulent kinetic energy through the interaction of the turbulent fluctuations with the mean-velocity gradient \(\pdv*{U(z)}{z}\).
The viscous dissipation term $\epsilon(z)$ involves the gradients of the fluctuating velocity.

In statistically homogeneous flows, production and dissipation are the only terms appearing in the turbulent kinetic energy balance.
In statistically \emph{inhomogeneous} flows, we also have spatial diffusion of turbulent kinetic energy $d(z)$.
The diffusion contains contributions associated with the turbulent fluctuations of the velocity and pressure (first two terms) and a contribution through viscous diffusion (the last term).
This viscous part of the diffusion is generally negligible compared to the contribution of the other two terms and will be dropped in the following.

The main question in the present investigation is how such inhomogeneous redistribution processes $d(z)$ affect the scaling of the kinetic energy spectrum $E(k,\vb*{x})$ in the inertial range of high Reynolds number turbulence.

\subsection{Fourier-analysis of inhomogeneous turbulence \label{subsec:Fourier-analysis}}

The use of energy spectra in general turbulent flows needs some justification.
In principle, Fourier modes are associated with infinite or periodic domains.
This property would exclude the use of spatial Fourier analysis of any realistic, non-periodic flow.
However, a closer look at the lengthscales involved in turbulent flows permits invoking an assumption of scale separation, allowing us to get around this problem.
Indeed, the theoretical basis for practical Fourier modelling of non-periodic turbulent flows can be found in various works~\citep[see][]{Jeandel1978_modeling, Yoshizawa1984_statistical, Bertoglio1987_a_simplified, Laporta1995_a_model, Besnard1996_spectral}.
In practice, to develop a spectral description of inhomogeneous flows, one needs to introduce a lengthscale $L$ characterising the inhomogeneity of the flow geometry.
Then, one can consider Fourier spectra associated with scales $r\sim k^{-1}$ small compared to $L$.

In the present investigation, we consider a spatially periodic flow without solid boundaries or obstacles to avoid most of these complications.
Furthermore, to derive corrections due to statistical inhomogeneity, we consider statistically stationary turbulence with a single inhomogeneous direction $z$.
An advantage of the present configuration, where only one inhomogeneous direction is present, is that we can compute energy spectra in planes perpendicular to the $z$-axis.
We thus define
\begin{equation}\label{eq:Ekperpz}
  E (k_\perp, z) \equiv \frac12 \int u_i (\vb*{k_\perp}, z) u_i^\ast (\vb*{k}_\perp, z) \dd{A (k_\perp)},
\end{equation}
where $A(k_{\perp})$ denotes a wavenumber-shell of radius $k_\perp$ in the $k_x, k_y$ plane.
The velocity field in~\eqref{eq:Ekperpz} is defined by the two-dimensional Fourier transform,
\begin{equation}
  u_i (\vb*{k}_\perp, z)
    \equiv \int e^{-\mathrm{i} \qty(k_x x + k_y y)} u_i(x, y, z) \dd{x} \dd{y}.
\end{equation}
The resulting energy spectrum \(E(k_\perp, z)\) is a function of a perpendicular wavenumber $k_\perp = \sqrt{k_x^2+k_y^2}$ and a vertical coordinate \(z\).
We note that if isotropy is restored in small scales, \(E(k_\perp, z)\) is expected to scale like the three-dimensional spectrum \(E(k, z)\) (see Appendix~\ref{app:Governing} for the definition).
In the following subsections in \S~\ref{sec:Derivation}, we will keep the notation $E(k, z)$ for the sake of generality, but it should be kept in mind that the scaling of $E(k, z)$ and $E(k_\perp, z)$ should be equivalent in statistically isotropic flow at large \(k\).


\subsection{Governing equation and modelling \label{subsec:Governing}}

The derivation in this subsection closely follows the rationale used to derive the instationary correction presented in~\citet{Bos2017_dissipation}.
This same methodology is here applied to the evolution-equation of the energy spectrum in inhomogeneous turbulence.

The kinetic energy spectrum is associated with the turbulent kinetic energy by the relation
\begin{equation}
  \int E(k, z) \dd{k} = K(z).
\end{equation}
The evolution equation for $E(k, z)$ is the multi-scale extension of equation~\eqref{eq:dKdt}.
This equation reads, for the case of a unidirectional mean flow $U(z)\vb*{e}_x$ as in~\eqref{eq:NSE},
\begin{equation}\label{eq:dEdt}
  \frac{\mathrm{D} E(k, z)}{\mathrm{D} t}
    =\underbrace{P(k, z)}_\textrm{production}
    -\underbrace{2\nu k^2 E(k, z)}_\textrm{dissipation}
    + \underbrace{T(k, z)}_\textrm{transfer}
    + \underbrace{D(k, z)}_\textrm{diffusion}.
\end{equation}
For self-consistency, we discuss the derivation of this equation in Appendix~\ref{app:Governing}.
Except for the viscous dissipation, all the terms in~\eqref{eq:dEdt} are unclosed.
In the following, we discuss the different physics and contributions to propose simple models for them.

Since the flow is statistically stationary and the mean flow is unidirectional, the material derivative on the left-hand side of~\eqref{eq:dEdt} is zero.
The first term on the RHS, $P(k, z)$, represents the terms directly proportional to the mean-velocity gradient.
It contains two contributions: the production of turbulent kinetic energy and a linear transfer term~\citep{Cambon1981_spectral, Briard2018_advanced}.
These terms are mainly important at large scales and become zero at points in space where the velocity gradient vanishes.
The order of magnitude of the production term can be estimated by~\citep{Tennekes1972_a_first},
\begin{equation}\label{eq:prod}
  P(k, z) \sim \qty(\pdv{U(z)}{z})^2 \tau(k, z) E(k, z),
\end{equation}
with the time scale $\tau(k, z) \sim \epsilon(z)^{-1/3} k^{-2/3}$ in the inertial range.
The integral of \(P(k, z)\) over wavenumbers yields $p(z)$ in~\eqref{eq:dKdt}.
Here, \(\epsilon(z)\) denotes the profile of the dissipation of kinetic energy through viscous stresses (see~\eqref{eq:epsilonz}) and is obtained by the integral of the second term on the RHS of~\eqref{eq:dEdt}.
At large Reynolds numbers, this term is significant only at large wavenumbers.
It is thus this term which is responsible for energy transfer between the mean velocity field \(U(z)\) and the turbulent kinetic energy.

The nonlinear transfer $T(k, z)$ represents the energy flux and is a redistributive term in scale space; thus, its integral over all wavenumbers yields zero.
The last term $D(k, z)$ represents the diffusion, or transport, through turbulent fluctuations and viscous diffusion.
Note that this term is zero in statistically homogeneous turbulence.
The term \(D(k, z)\) is also a redistribution term like $T(k, z)$, but in physical space.
Its integral over wavenumbers corresponds to $d(z)$ in~\eqref{eq:dKdt}.

\begin{figure}
  \centering
  \includegraphics[width=0.6\textwidth]{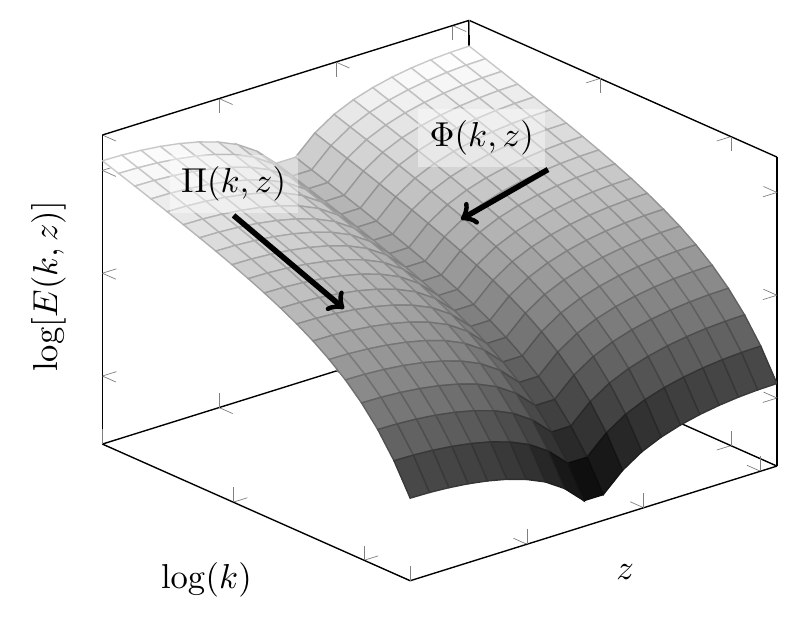}
  \caption{
    A schematic of the energy spectrum in \((k,z)\) coordinates.
    Two arrows denote the direction of energy fluxes in wavenumber and physical space, respectively.
  }
  \label{Fig:flux_kz}
\end{figure}

Both $T(k, z)$ and $D(k, z)$ are a function of triple correlations between Fourier modes at different wavelengths.
There is no exact expression of these quantities as a closed function of the kinetic energy spectrum \(E(k, z)\).
At this moment, we will therefore introduce modelling assumptions.
Sophisticated models exist for inhomogeneous spectral dynamics, based on the Test Field Model~\citep{Kraichnan1971_almost} or the Eddy-Damped Quasi-Normal Approximation~\citep{Laporta1995_a_model, Parpais1996_a_spectral}.
However, the resulting closures are quite complicated and do not allow a straightforward analytical perturbation treatment.
Therefore, our approach uses simple models that reproduce their main physical features: the redistribution of energy in scale space for $T(k, z)$ and in physical space for $D(k, z)$, respectively.
We use diffusion approximations for both terms,
\begin{empheq}{align}
  T(k, z) &= -\pdv{k} \Pi(k, z),
    \label{eq:Tdpi}\\
  D(k, z) &= -\pdv{z} \Phi(k, z),
    \label{eq:Ddpi}
\end{empheq}
where $\Pi(k, z)$ and $\Phi(k, z)$ are turbulent fluxes in wavenumber and physical space, respectively.
Figure~\ref{Fig:flux_kz} schematically depicts these two fluxes in \((k,z)\) space.
In the absence of inhomogeneity, the flux \(\Phi(k, z)\) is zero.
In the inhomogeneous case, the presence of this flux will affect the kinetic energy spectrum \(E(k, z)\).

We model both fluxes using a gradient-diffusion approximation,
\begin{equation}\label{eq:model_Pi}
  \Pi(k, z) = -\rho(k, z) \pdv{\qty(k^{-2}E(k, z))}{k}
\end{equation}
with $\rho(k, z) \sim k^{11/2}E(k, z)^{1/2}$ being a turbulent energy diffusion in Fourier space, and
\begin{equation}\label{eq:ddze}
  \Phi(k, z) = -\mu(z) \pdv{E(k, z)}{z},
\end{equation}
where $\mu(z)$ is a turbulent diffusivity in real space (see~\eqref{eq:mu_z}).
We have effectively decoupled (and simplified) the transfer terms in scale and physical space.
Indeed, both \(\Phi(k, z)\) and \(\Pi(k, z)\) are determined by the same triple velocity and velocity-pressure correlations (see~\eqref{eq:Psi_ij}).
Decomposing the physical space-scale space flux is a major assumption which seems necessary to obtain an analytically tractable model of energy transfer in inhomogeneous turbulence.
The model for $\Pi(k, z)$~\eqref{eq:model_Pi} is known as the Leith model~\citep{Leith1967_diffusion, Rubinstein2022_reassessment}.
This model tends to homogenise the kinetic energy in spectral space towards equipartition among wave vectors, corresponding to an energy spectrum proportional to $k^2$.
The gradient-diffusion model for the diffusion~\eqref{eq:ddze} tends to homogenise the energy distribution in physical space and is used in~\cite{Besnard1996_spectral, Touil2002_the_decay, Cadiou2004_a_two-scale}, for instance.

Eddy viscosity models are obviously simplified representations of the real transfer terms.
For instance, see~\citet[\S~10]{Pope2000_turbulent} for extensive discussions.
However, we think that this kind of modelling is a useful first step before turning to more sophisticated modelling approaches.

\subsection{Linear perturbation analysis and scaling predictions \label{subsec:Linear}}

Our goal is to derive a prediction for inertial range scaling at large Reynolds numbers in the limit of weak inhomogeneity, where the influence of inhomogeneity can be treated as a perturbation.
In the following, the leading order contributions and perturbations are indicated by a subscript $0$ and $1$, respectively.
We define an inertial range \(L^{-1} \ll k \ll \eta^{-1}\) with the length $L$ representing the typical length of the largest and energy-containing scales of the flow.
Furthermore, in our description, it is associated with the longest wavelength in our flow domain and is chosen constant.
We will define this length scale more precisely later, in \S~\ref{subsec:Case}.

We now define the equilibrium about which we expand the equations.
To do so, we consider the decomposition
\begin{equation}
  E(k, z)
    = E_0(k, z) + E_1(k, z)
  \label{eq:E_decomposition}
\end{equation}
with \(\abs{E_1} \ll \abs{E_0}\).
The other quantities, such as \(\Pi(k, z)\) and \(\Phi(k, z)\), are decomposed in the same manner.
We recall here that in addition to these two contributions to the energy spectrum, the flow also contains the time-averaged velocity profile, which consists of a single wave vector in the \(z\)-direction in the present case~\eqref{eq:K_Uz}.
This mean flow is not present in the inertial range, on which we will focus in the following.
Therefore, in the remainder of this section, we can focus on the contributions \(E_0(k, z)\) and \(E_1(k, z)\).

For very high Reynolds numbers in the limit of vanishing inhomogeneity, we assume that the equilibrium contributions to the kinetic energy balance~\eqref{eq:dEdt} do not depend on the inhomogeneous turbulent diffusion $D(k, z)$.
By integrating the balance between the transfer and dissipation terms in~\eqref{eq:dEdt} from $k$ to $\infty$, we find
\begin{equation}
  \int_k^\infty T(p, z) \dd{p}
    = \int_k^\infty 2 \nu p^2 E(p, z) \dd{p}
\end{equation}
or, using expression~\eqref{eq:Tdpi} and the equilibrium/nonequilibrium decomposition,
\begin{equation}
  \Pi_0(k, z) =\epsilon(z).
\end{equation}
Indeed, this corresponds to the equilibrium between the energy flux and the energy dissipation rate, essential to the inertial range description of~\citet{Kolmogorov1941_the_local}.
The constant flux solution of the Leith model is consistent with this framework and is given by
\begin{equation}\label{eq:E0}
  E_0(k, z) \sim \epsilon(z)^{2/3} k^{-5/3}.
\end{equation}
This expression defines our equilibrium solution.
We now assess the influence of the inhomogeneity of $\epsilon(z)$ on this scaling as a perturbation.

In the following, we consider the terms in the balance equation~\eqref{eq:dEdt} for the nonequilibrium contributions.
The order of magnitude of the production term~\eqref{eq:prod} and the diffusion-gradient modelling with the flux~\eqref{eq:ddze} leads us to deduce that $D(k, z)\gg P(k, z)$ at $k \gg L^{-1}$.
Therefore, the first-order perturbation to the equilibrium scaling in the inertial range is due to the inhomogeneous diffusion \(D(k, z)\).  
Then, in the inertial range, we have
\begin{equation}
  T(k, z) = -D(k, z),
\end{equation}
and
\begin{equation}\label{eq:dkpi1}
  -\pdv{k}\Pi_1(k, z) = \pdv{z} \Phi_0 (k, z),
\end{equation}
since \(\pdv*{\Pi_0(k, z)}{k} = 0\).
Thus, the \emph{first}-order correction of the nonlinear transfer balances the \emph{zeroth}-order contribution of the inhomogeneous diffusion.
The first-order perturbation to the nonlinear flux $\Pi_1(k, z)$ is evaluated as~\citep{Rubinstein2005_self-similar}
\begin{equation}\label{eq:pi1}
  \Pi_1(k, z) = E_1(k, z) \eval{\fdv{\Pi}{E}}_{E_0},
\end{equation}
where $\eval{\fdv*{\Pi}{E}}_{E_0}$ is the Fr\'echet derivative of the total flux $\Pi$ evaluated at $E(k, z) = E_0(k, z)$.
In the inertial range, assuming \(E_1\) to scale as a power law, this yields the scaling,
\begin{equation}\label{eq:pie1e0}
  \Pi_1(k, z) \sim \epsilon(z) \frac{E_1(k, z)}{E_0(k, z)}.
\end{equation}
Note that we obtain~\eqref{eq:pie1e0} not only for the Leith model, but also for most of the other classical closures such as the Kovaznay and Heisenberg model~\citep{Rubinstein2022_reassessment}.
Integrating~\eqref{eq:dkpi1} from $k$ to $\infty$, we have
\begin{equation}\label{eq:dkpi2}
  \Pi_1 (k, z) = \pdv{z} \int_k^\infty \Phi_0 (k, z) \dd{k}.
\end{equation}
By combining this with~\eqref{eq:ddze} and~\eqref{eq:pie1e0}, we obtain
\begin{equation}
  E_1 (k, z) \sim - \frac{E_0(k, z)}{\epsilon(z)} \pdv{z} \qty( \mu(z) \pdv{\int_k^\infty E_0 (p, z) \dd{p}}{z} ).
\end{equation}
Substituting~\eqref{eq:E0}, the above expression gives
\begin{equation}
  E_1(k, z) \sim -\mu(z) \epsilon(z)^{1/3} k^{-7/3} \qty[
    \frac23 \frac{\epsilon_{zz}(z)}{\epsilon(z)}
    + \frac23 \frac{\mu_z(z)}{\mu(z)} \frac{\epsilon_z(z)}{\epsilon(z)}
    - \frac29 \qty(\frac{\epsilon_z(z)}{\epsilon(z)})
  ],
\end{equation}
where the subscripts denote derivatives with respect to \(z\), for example, \(\epsilon_{zz} = \partial^2 \epsilon(z) / \partial z^2\).
We will model the unknown eddy diffusivity in its simplest way,
\begin{equation}\label{eq:mu_z}
  \mu(z) \sim  L^{4/3} \epsilon(z)^{1/3}.
\end{equation}
Doing so, we obtain
\begin{equation}\label{eq:E1_spatial}
  E_1(k, z) \sim -\frac{\epsilon_{zz}(z) L^{4/3}}{\epsilon(z)^{1/3}} k^{-7/3}.
\end{equation}
Note that although all the terms involving \(\epsilon_z\) and \(\mu_z\) vanish exactly for the current definition of \(\mu(z)\) in~\eqref{eq:mu_z}, this might not be the case for arbitrary choices of \(\mu(z)\).



\subsection{Case of a sinusoidal dissipation profile \label{subsec:Case}}

The comparison of expressions~\eqref{eq:E0} and~\eqref{eq:E1_spatial} indicates that the inhomogeneous contribution (\(\propto k^{-7/3}\)) is subdominant compared to the equilibrium energy spectrum (\(\propto k^{-5/3}\)) at large wavenumbers.
Furthermore, the expression is proportional to the second spatial derivative of the dissipation rate \(\epsilon_{zz}(z)\) and can thus be both positive and negative.
Let us illustrate the implication of this expression by considering a large-scale inhomogeneity characterised by a cosine function with a characteristic wavelength of order $L$,
\begin{equation}\label{eq:epsilon_fitting}
  \epsilon(z) = \expval{\epsilon} + \tilde{\epsilon} \cos(z/L),
\end{equation}
with $\expval{\epsilon} \gg \tilde{\epsilon}$.
We consider \(L\), first introduced in \S~\ref{subsec:Linear}, to be of the order of and proportional to the characteristic large-scale length of the flow.
Substituting this expression for \(\epsilon(z)\) in~\eqref{eq:E1_spatial}, we find
\begin{equation}\label{eq:E1ca2}
  E_1(k, z) = E_1^X(k)\cos(z/L)
\end{equation}
with
\begin{equation}\label{eq:E1ca3}
  E_1^X(k) = C_A \tilde\epsilon \expval{\epsilon}^{-1/3}L^{-2/3} k^{-7/3},
\end{equation}
where the superscript \(X\) indicates the perturbations due to inhomogeneity.

Let us now assume that both the equilibrium spectrum $E_0(k, z)$ and $E_1(k, z)$ extend from $k=L^{-1}$ to $\infty$.
Integrating the spectra in this range, we find that
\begin{equation}\label{eq:K0_sim_eps23}
  K_0(z) \sim L^{2/3} \epsilon(z)^{2/3}
\end{equation}
and
\begin{equation}\label{eq:K1_sim_epszz_K0}
  K_1(z) \sim -\frac{\epsilon_{zz}(z) L^2}{\expval{\epsilon}} K_0(z).
\end{equation}
Comparing these last two expressions illustrates that the formal expansion parameter in our system is
\begin{equation}
  \gamma = \frac{\epsilon_{zz}(z) L^2}{\expval{\epsilon}}.
\end{equation}

The main analytical results of the present investigation [\eqref{eq:E1ca3}--\eqref{eq:K1_sim_epszz_K0}] are obtained by phenomenological modelling based on gradient-diffusion assumptions of nonlinear transfer in both physical and scale space.
The models and their consequences are, at best, crude approximations of the intricate nonlinear interactions in the actual flow.
Therefore, The resulting expressions need verification by experiments or direct numerical simulations.

\section{Assessment of the inhomogeneous scaling \label{sec:Assessment}}

\subsection{Numerical set-up \label{subsec:Numerical}}

In order to verify the theoretical predictions, in particular expression~\eqref{eq:E1ca3}, we carry out DNS of three-dimensional Kolmogorov flow in a triple-periodic box.
Such flow has the convenient properties of being statistically inhomogeneous in one direction and free of solid boundaries.
Furthermore, its properties have been widely investigated numerically~\citep{Borue1996_numerical, Musacchio2014_turbulent, Wu2021_a_quadratic}.

\begin{table}
  \centering
  \def~{\hphantom{0}}
  \begin{tabular}{cccccc}
    $N$ & $\nu$ & $u'$ & $\lambda$ & $\Rey_\lambda$ & $T_\textrm{total} / T$ \\[3pt]
    $128$ & $0.07$ & $1.31$ & $0.371$ & $69.6$ & $959$ \\  
    $256$ & $0.028$ & $1.35$ & $0.233$ & $113$ & $645$ \\  
    $512$ & $0.01$ & $1.33$ & $0.138$ & $184$ & $170$  
  \end{tabular}
  \caption{
    DNS parameters and statistical quantities.
    The resolution \(N\) and kinematic viscosity \(\nu\) are the control parameters.
    The remaining statistical quantities are:
    the fluctuating isotropic RMS velocity \(u' \equiv \sqrt{2K'/3}\) where energy of the temporal fluctuating velocity \(K' \equiv \expval{u'_i u'_i}_{\vb*{x}, t} / 2 \) and \(u'_i (\vb*{x}, t) \equiv u_i (\vb*{x}, t) - \expval{u_i}_t(\vb*{x})\);
    the Taylor microscale \(\lambda \equiv u' \sqrt{15 \nu / \epsilon}\) where the energy dissipation rate is evaluated by \(\epsilon = \nu \expval{\omega_i \omega_i}_{\vb*{x}, t}\);
    the Taylor-length Reynolds number \(\Rey_\lambda \equiv u' \lambda / \nu\);
    the integral time scale \(T \equiv L/u'\) with \(L = k_f^{-1} = 1\);
    the simulation time in the statistically steady state \(T_\textrm{total}\) as a function of \(T\).
  }
  \label{tab:DNS_setup}
\end{table}
The dynamics of the Kolmogorov flow in the present investigation are governed by~\eqref{eq:NSE} with \(f(z) = \sin(k_f z)\).
The numerical domain is a cube of size $2\upi$.
These choices imply that the forcing wavelength is equal to the width of the cubic domain, and we set \(k_f = L^{-1} = 1\).
Simulations are carried out using a standard pseudo-spectral solver~\citep{Delache2014_scale} with a third-order Adams-Bashfort time-integration scheme.
The details of the simulations are reported in table~\ref{tab:DNS_setup}.
Since we focus on the effect of inhomogeneity, we attempt to obtain statistics in a steady state over a long-enough time interval to allow the effects of the temporal variations to become as small as possible (see the last column in table~\ref{tab:DNS_setup}).

\subsection{Visualisation and dissipation profile \label{subsec:Visualisation}}

\begin{figure}
  \centering
  \includegraphics[width=0.8\textwidth]{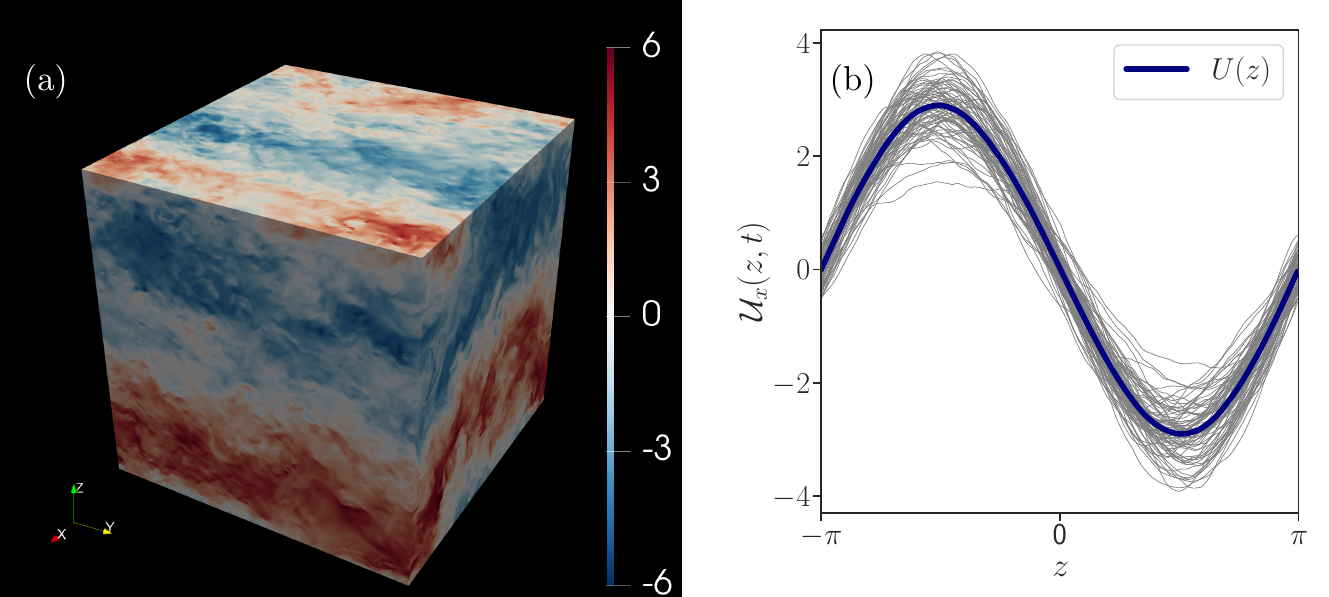}
  \caption{
    (a) Instantaneous distributions of \(\mathcal{U}_x(\vb*{x}, t)\) at \(\Rey_\lambda=184\).
    Blue (red) corresponds to the negative (positive) value of \(\mathcal{U}_x\).
    (b) Instantaneous profiles of \(\mathcal{U}_x(z, t) = \expval{\mathcal{U}_x(\vb*{x}, t)}_{\perp}\) in grey.
    Time-averaged profile \(U(z) = \expval{\mathcal{U}_x(\vb*{x}, t)}_{\perp, t}\) is indicated by a thick line.
  }
  \label{Fig:visu}
\end{figure}

In the following, we will discuss the simulation at the highest considered Reynolds number $\Rey_\lambda=184$.
A flow visualisation is shown in figure~\ref{Fig:visu}~(a) with the \(x\)-component of the velocity field \(\mathcal{U}_x (\vb*{x}, t)\).
The influence of the large-scale mean flow, proportional to the sinusoidal forcing along the \(z\) axis, is distinguishable.
Figure~\ref{Fig:visu}~(b) shows the instantaneous profile of \(\mathcal{U}_x (z, t) = \expval{\mathcal{U}_x (\vb*{x}, t)}_{\perp}\).
The single curve corresponds to the horizontal average of a snapshot, as shown in figure~\ref{Fig:visu}~(a).
Its time average, \(U(z) = \expval{\mathcal{U}_x (\vb*{x}, t)}_{\perp, t}\), is also shown in figure~\ref{Fig:visu}~(b) with a smooth sinusoidal profile.

\begin{figure}
  \centering
  \includegraphics[width=0.8\textwidth]{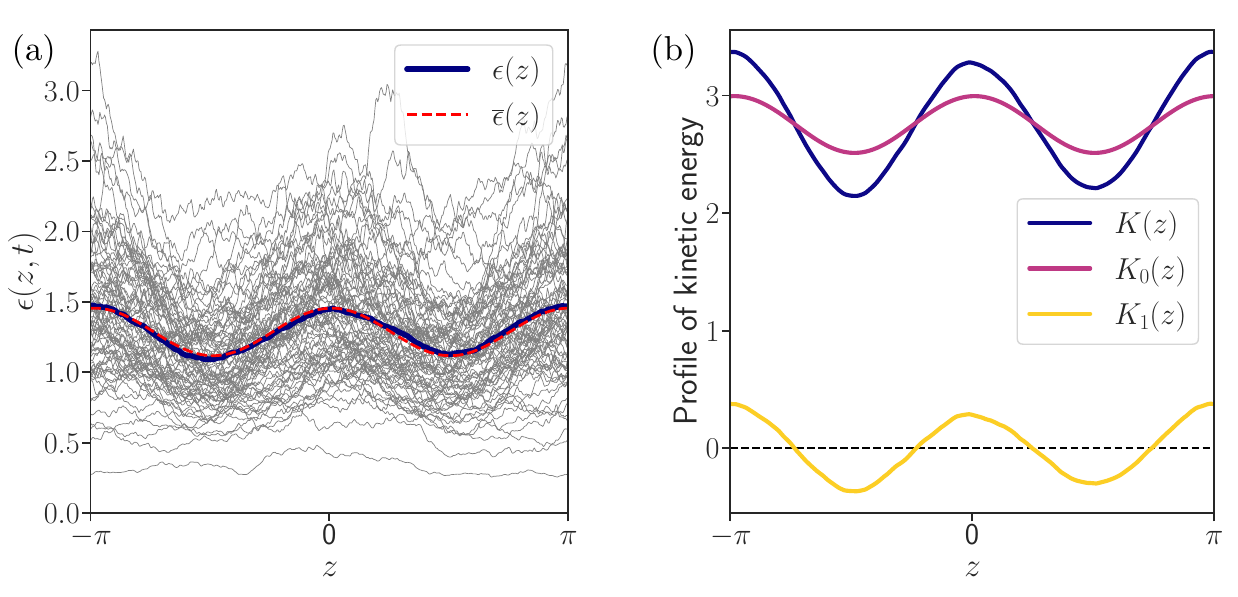}
  \caption{
    (a) Instantaneous profile of \(\epsilon(z, t) = \expval{\epsilon(\vb*{x}, t)}_{\perp}\).
    Time-averaged profile \(\epsilon(z) = \expval{\epsilon(\vb*{x}, t)}_{\perp, t}\) is also shown.
    The red dashed line denotes \(\overline{\epsilon}(z)\), a sinusoidal fitting of \(\epsilon(z)\) by~\eqref{eq:epsilon_fitting}.
    (b) Time-averaged profile of kinetic energy with fluctuating velocity \(K(z)\) and its equilibrium \(K_0(z)\) and nonequilibrium \(K_1(z)\) contributions.
    See the main text and Appendix~\ref{app:Normalisation} for the definition.
  }
  \label{Fig:eps_K_profile}
\end{figure}

In figure~\ref{Fig:eps_K_profile}~(a), the instantaneous profile of the energy dissipation rate \(\epsilon(z, t) = \expval{\epsilon(\vb*{x}, t)}_{\perp}\) is shown along with its time average \(\epsilon(z) = \expval{\epsilon(\vb*{x}, t)}_{\perp, t}\).
The instantaneous profile shows large fluctuations in comparison to the velocity profile (figure~\ref{Fig:visu}~(b)).
Its time average, in contrast, shows a smooth sinusoidal profile.
This property allows us to use the approximations in~\S~\ref{subsec:Case}.
As expected, the dissipation peaks at values where the mean velocity gradient is strongest (at \(z = 0\) and \(\pm\pi\)).
For numerical convenience, we perform a sinusoidal fitting \(\overline{\epsilon}(z)\) introduced in~\eqref{eq:epsilon_fitting}.
This profile is also shown in figure~\ref{Fig:eps_K_profile}~(a).

Figure~\ref{Fig:eps_K_profile}~(b) shows the kinetic energy profile of the fluctuating velocity field.
The fluctuating energy profile is defined by \(K(z, t) = \mathcal{K}(z, t) - K_U(z)\), where the total energy is \(\mathcal{K}(z, t) = \mathcal{U}_i (z, t) \mathcal{U}_i (z, t) / 2\) and the mean flow energy is \(K_U(z) = U(z)^2 / 2\).
We consider the decomposition, see~\eqref{eq:K0_sim_eps23}--\eqref{eq:K1_sim_epszz_K0} and Appendix~\ref{app:Normalisation},
\begin{equation}
  K(z, t) = K_0(z, t) + K_1(z, t).
  \label{eq:K_decomposition}
\end{equation}
In figure~\ref{Fig:eps_K_profile}~(b), we observe that the equilibrium \(K_0(z) = \expval{K_0(z, t)}_t\) and the nonequilibrium \(K_1(z) = \expval{K_1(z, t)}_t\) profiles share the same phase, consistent with the prediction that the spectrum \(E_1(k, z)\) is proportional to \(-\pdv*[2]{\epsilon(z)}{z}\).

\subsection{Equilibrium and nonequilibrium spectra \label{subsec:Equilibrium}}

\begin{figure}
  \centering
  \includegraphics[width=0.6\textwidth]{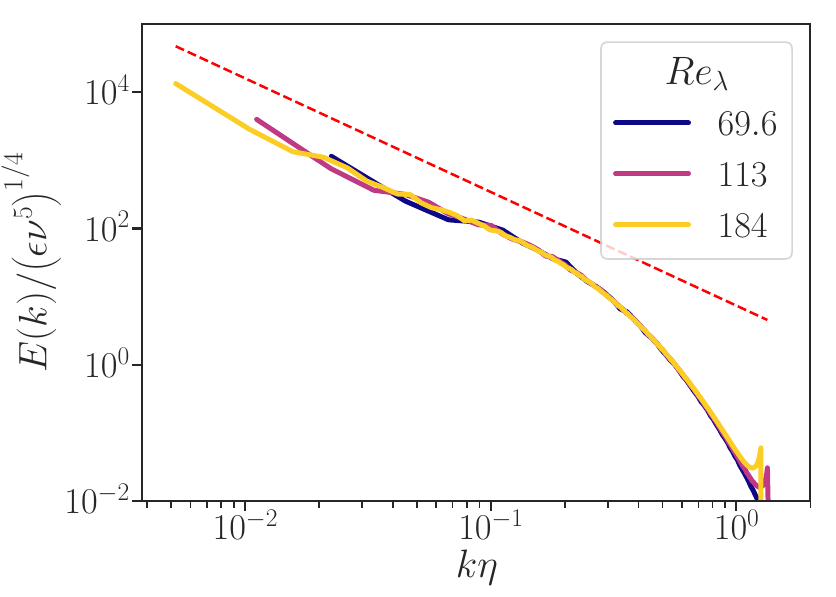}
  \caption{
    Time-averaged three-dimensional isotropic energy spectrum \(E(k) = \expval{E(k, t)}_t\), normalised by Kolmogorov variables. Results are shown at $\Rey_\lambda=69.6, 113$ and $184$ (see table~\ref{tab:DNS_setup}).
    The red dashed line denotes the \(k^{-5/3}\) scaling for reference.
  }
  \label{Fig:Ek}
\end{figure}

Figure~\ref{Fig:Ek} shows the isotropic energy spectrum \(E(k, t)\) (see Appendix~\ref{app:Governing} for the definition) at three different Taylor-length Reynolds numbers.
For simplicity, we denote its time-average by \(E(k) = \expval{E(k, t)}_t\).
Normalisation using \(\nu\) and \(\epsilon = \expval{\epsilon(\vb*{x}, t)}_{\vb*{x}, t}\) allows an excellent collapse for large values of $k$.

\begin{figure}
  \centering
  \includegraphics[width=\textwidth]{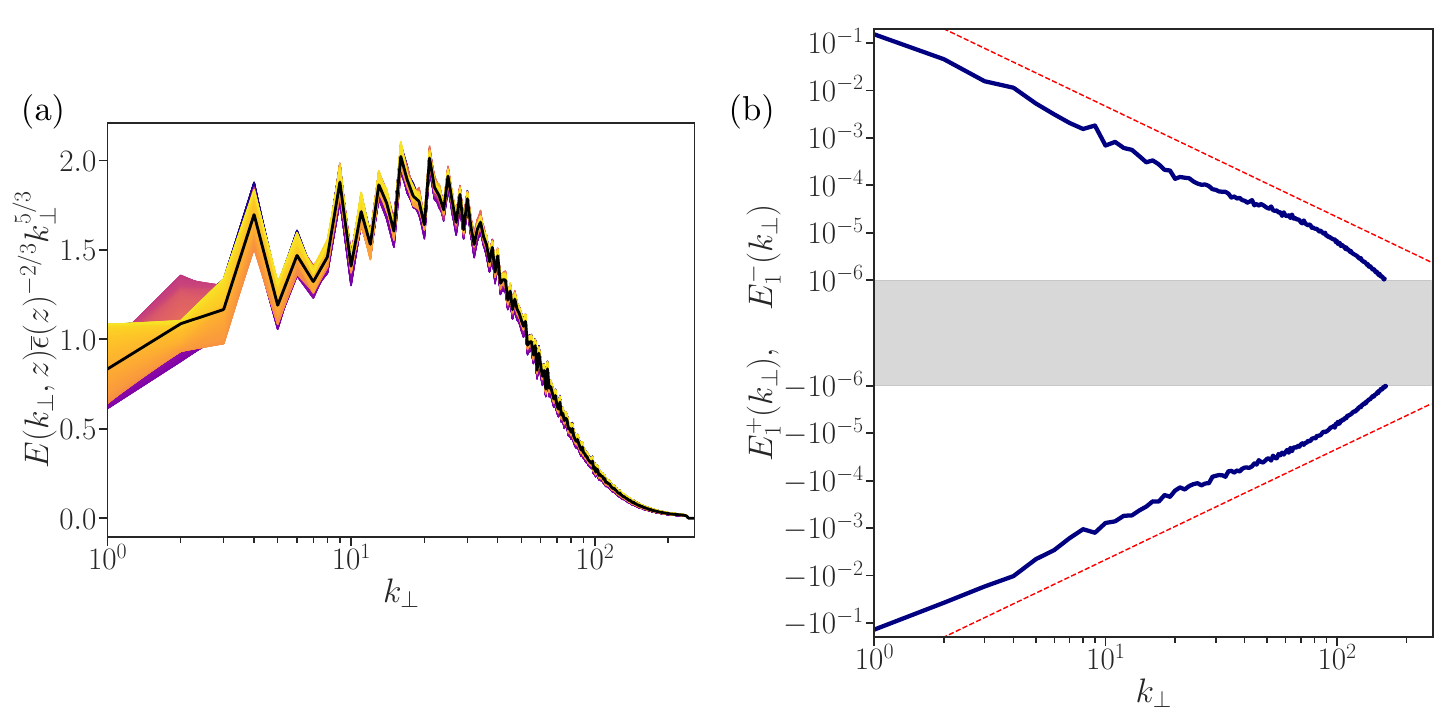}
  \caption{
    (a) Nondimensionalised two-dimensional energy spectrum.
    Note that \(E(k_\perp, z) = \expval{E(k_\perp, z, t)}_t\).
    Dark (light) colour represents the small (large) value of \(z\) coordinate.
    The thick black line denotes~\eqref{eq:F_epsilon_L}, the average over \(z\) coordinate.
    (b)~Time-averaged nonequilibrium energy spectrum with specific signs: \(E_1^+ (k_\perp) = \expval{E_1(k_\perp, z) > 0}_z\) and \(E_1^- (k_\perp) = \expval{E_1(k_\perp, t) < 0}_z\).
    Red dashed lines denote the \(k_\perp^{-7/3}\) slope.
  }
  \label{Fig:E1}
\end{figure}

Next, we assess energy spectra in statistically homogeneous planes perpendicular to the \(z\) axis, as defined in~\eqref{eq:Ekperpz}.
In the following, we analyse the time-averaged inhomogeneous energy spectrum \(E(k_\perp, z) = \expval{E(k_\perp, z, t)}_t\) in a statistically steady state (see table~\ref{tab:DNS_setup}).
Figure~\ref{Fig:E1}~(a) shows \(E(k_\perp, z)\) nondimensionalised by \(\overline{\epsilon}(z)^{2/3} L^{2/3}\).
The fluctuations at small scales are small, and variations are barely visible.

Then, we assume that the energy spectrum in the inertial range can be written, in a general form,
\begin{equation}\label{eq:E0_normalisation}
  E_0(k_\perp, z) \sim \epsilon(z)^{2/3} k_\perp^{-5/3} f_L[k_\perp L] f_\eta[k_\perp \eta(z)],
\end{equation}
where \(f_L\) and \(f_\eta\) determine the shape at small and large \(k\), respectively.
Therefore, there are two distinct choices to collapse the spectra.
See Appendix~\ref{app:Normalisation} for the detail of two normalisation procedures.
In this study, we employ the large-scale normalisation and evaluate
\begin{equation}
  f_L[k_\perp L] \equiv \expval{E(k_\perp, z) \overline{\epsilon}(z)^{-2/3} k_\perp^{5/3}}_z,
  \label{eq:F_epsilon_L}
\end{equation}
as shown in figure~\ref{Fig:E1}~(a).
Note that this expression is valid for \(k_\perp \eta \ll 1\) where \(f_\eta (k_\perp \eta)\) tends to a constant value.
Then, the equilibrium spectrum can be defined as
\begin{equation}
  E_0 (k_\perp, z) \equiv \overline{\epsilon}(z)^{2/3} f_L[k_\perp L] k_\perp^{-5/3}.
  \label{eq:E0_epsilon_F}
\end{equation}

Now, we can evaluate the nonequilibrium spectrum by \(E_1(k_\perp, z) \equiv E(k_\perp, z) - E_0(k_\perp, z)\).
Note that
i) this quantity is defined by the time-averaged spectra and
ii) since this quantity can be regarded as a perturbation of \(E(k_\perp, z)\) around \(E_0(k_\perp, z)\), it can be both positive and negative.
Figure~\ref{Fig:E1}~(b) shows the \(z\)-average of \(E(k_\perp, z)\) for specific signs.
Similar plots are shown in figure~10 of~\cite{Horiuti2011_multimode} and figure~2 of~\cite{Horiuti2013_nonequilibrium}.
The scaling is consistent with the one derived in \S~\ref{subsec:Linear},
\begin{equation}
  \expval{\abs{E_1(k_\perp, z)}}_z \propto k_\perp^{-7/3}.
\end{equation}

\begin{figure}
  \centering
  \includegraphics[width=\textwidth]{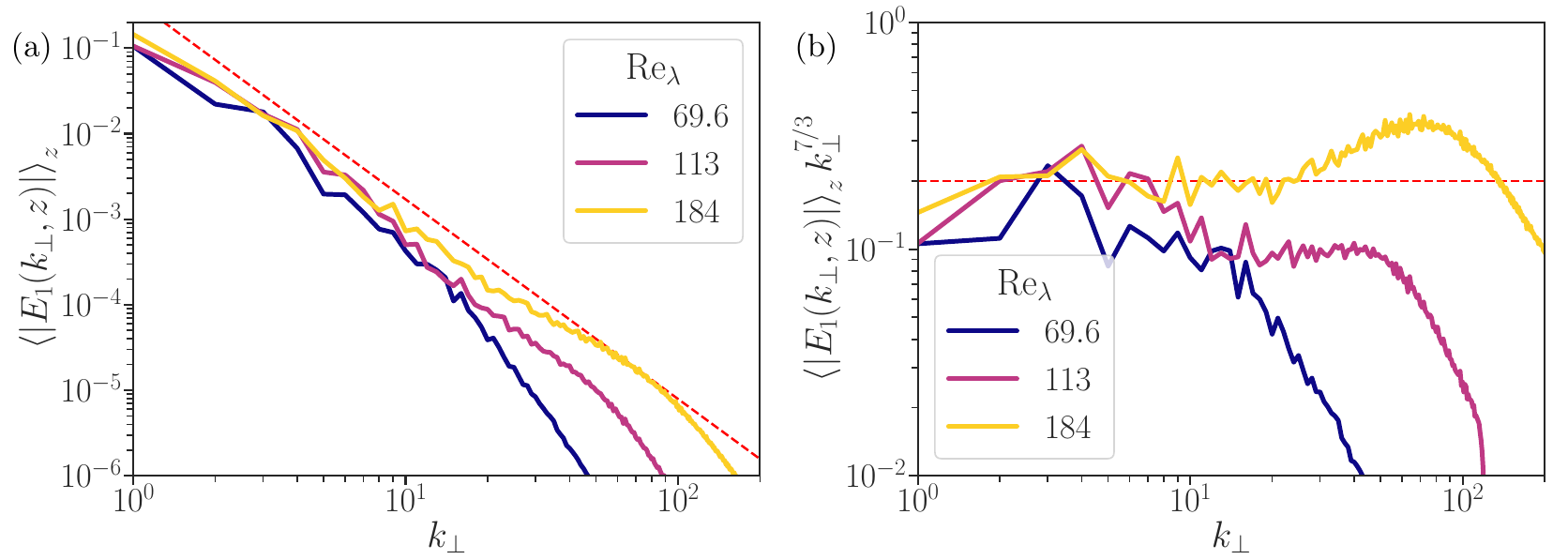}
  \caption{
    (a)~Absolute value of the time-averaged nonequilibrium energy spectrum \(\abs{E_1(k_\perp, z)} = \abs{\expval{E_1(k_\perp, z, t)}_t}\) for three values of the Taylor-length Reynolds numbers.
    The red dashed line represents the \(k_\perp^{-7/3}\) scaling.
    (b)~Compensated spectrum of the panel~(a).
    The red dashed line denotes the compensated \(k_\perp^{-7/3}\) scaling.
  }
  \label{Fig:E1_Re}
\end{figure}

Figure~\ref{Fig:E1_Re}~(a) compares \(\expval{\abs{E_1(k_\perp, z)}}_z\) for three different Reynolds numbers as in figure~\ref{Fig:Ek}.
For smaller values of \(\Rey_\lambda\), the spectrum exhibits steeper scaling than \(k_\perp^{-7/3}\).
At larger \(\Rey_\lambda\), the slope approaches the \(k_\perp^{-7/3}\) scaling.
At the same time, the spectrum in the higher \(k_\perp\) range exhibits a bump associated with shallower scaling than \(k_\perp^{-7/3}\).

We plot the compensated spectra in figure~\ref{Fig:E1_Re}~(b).
Although the scaling range extends for less than a decade, the emergence of the \(k_\perp^{-7/3}\) scaling range is well captured using this normalisation.
In Appendix~\ref{app:Normalisation}, we confirm that the bump in the compensated spectra is due to our choice of the nondimensional function~\eqref{eq:E0_normalisation}.

\section{Conclusion and discussion \label{sec:Conclusion}}

The numerical simulations in the previous section support our prediction,
\begin{equation}
  E(k, z)
    = C_K \epsilon(z)^{2/3} k^{-5/3}
    - C_A \frac{\epsilon_{zz}(z) L^{4/3}}{\epsilon(z)^{1/3}} k^{-7/3},
\end{equation}
of the energy spectrum for turbulence with inhomogeneity in the \(z\) direction.
This scaling quantifies the influence of spatial inhomogeneity in wavenumber space.
In particular, the special case where the dissipation fluctuates as a sinusoidal function around a mean value, discussed in~\S\ref{subsec:Case}, gives us a useful estimate of the influence of inhomogeneity~\eqref{eq:E1ca2}--\eqref{eq:E1ca3}.
Indeed, introducing an average dissipation $\expval{\epsilon}$ and smooth spatial fluctuations \(\tilde{\epsilon}\) around \(\expval{\epsilon}\), so that $\epsilon_{zz}/\tilde\epsilon\sim L^{-2}$, we obtain that
\begin{equation}
  \frac{E^X_1(k, z)}{E_0(k, z)} \sim \frac{\tilde{\epsilon}}{\expval{\epsilon}}(kL)^{-2/3}.
\end{equation}
This expression shows that the influence of large-scale inhomogeneity is negligible for
\begin{equation}
  k \gg L^{-1}\qty(\frac{\tilde{\epsilon}}{\expval{\epsilon}})^{3/2}.
\end{equation}
Therefore, if this requirement is fulfilled in a statistically stationary flow, far enough away from walls, Kolmogorov's equilibrium spectrum is expected to be dominant compared to the contributions associated with spatial inhomogeneity.

\appendix
\section{Governing equation of the inhomogeneous energy spectrum \label{app:Governing}}

In this Appendix, we define the spectrum and the governing equations for the energy spectrum tensor in inhomogeneous flow (see~\eqref{eq:dEdt}).
The generalised spectrum \(E(k, \vb*{x}) = E_{ii} (k, \vb*{x})\) is defined by
\begin{equation}
  E(k, \vb*{x}) \equiv \frac12 \iint e^{-\mathrm{i} \vb*{k} \vdot \vb*{r}} \expval{
    u_i \qty(\vb*{x} + \frac{\vb*{r}}{2}) u_i \qty(\vb*{x} - \frac{\vb*{r}}{2})
  } \dd{\vb*{r}} \dd{\Omega_k},
\end{equation}
where \(\int \dd{\Omega_k}\) denotes the integral over spherical shells of radius \(k\).
The brackets in this Appendix section denote the ensemble average.
For statistically homogeneous turbulence, this definition is equivalent to the expression
\begin{equation}
  E(k) \equiv \int \frac12 \expval{u_i (\vb*{k}) u_i^\ast (\vb*{k})} \dd{\Omega_k}.
\end{equation}
The evolution equation for \(E_{ij} (k, \vb*{x}, t)\) formally reads
\begin{equation}
  \pdv{E_{ij} (k, \vb*{x}, t)}{t}
    = \frac12 \int \qty[ \int e^{-\mathrm{i} \vb*{k} \vdot \vb*{r}} \Psi_{ij} (\vb*{x} + \vb*{r}/2, \vb*{x} - \vb*{r}/2, t) \dd{\vb*{r}} ] \dd{\Omega_k}.
\end{equation}
For the tensor on the RHS, we have
\begin{empheq}{align}
  \Psi_{ij} (\vb*{x}_1, \vb*{x}_2) =
    & \nu \qty(\laplacian_1 + \laplacian_2) R_{ij} (\vb*{x}_1, \vb*{x}_2) \nonumber \\
    &- \Bigl[ \pdv{x_{1n}} U_n (\vb*{x}_1) R_{ij} (\vb*{x}_1,\vb*{x}_2)
    + \pdv{x_{2n}} U_n (\vb*{x}_2) R_{ij} (\vb*{x}_1,\vb*{x}_2)\nonumber\\
    &+ \pdv{x_{1n}} U_i (\vb*{x}_1) R_{nj} (\vb*{x}_1,\vb*{x}_2)
    + \pdv{x_{2n}} U_j (\vb*{x}_2) R_{ni} (\vb*{x}_1,\vb*{x}_2)\nonumber\\
    &+ \pdv{x_{1i}} \expval{ \mathcal{P}(\vb*{x}_1) u_j(\vb*{x}_2) }
    + \pdv{x_{2j}} \expval{ \mathcal{P}(\vb*{x}_2) u_i(\vb*{x}_1) } \nonumber\\
    &+ \pdv{x_{1n}} \expval{ u_i(\vb*{x}_1) u_n(\vb*{x}_1) u_j(\vb*{x}_2) }
    + \pdv{x_{2n}} \expval{ u_j(\vb*{x}_2) u_n(\vb*{x}_2) u_i(\vb*{x}_1) } \Bigr].
  \label{eq:Psi_ij}
\end{empheq}
In this expression and the following, the argument \(t\) for time is omitted for visibility.
The two-point velocity tensor is defined by
\begin{equation}
  R_{ij} (\vb*{x}_1, \vb*{x}_2)
    \equiv \expval{u_i (\vb*{x}_1) u_j (\vb*{x}_2)},
\end{equation}
and the equation~\eqref{eq:Psi_ij} is completed by incompressibility conditions for the mean field and the fluctuations.

Both inhomogeneous turbulence diffusion and spectral transfer are associated with the last two lines of expression~\eqref{eq:Psi_ij}.
An assumption of weak inhomogeneity must be invoked to dissociate them to obtain a closed expression.
Even retaining only the leading order terms in an expansion about inhomogeneity, the resulting equations become quite cumbersome (see~\cite{Laporta1995_etude, Besnard1996_spectral}).

Subsequently, the different terms in~\eqref{eq:Psi_ij} need to be modelled to close the triple correlations.
We will not proceed in this direction and will directly model them by their physical effects.
See~\eqref{eq:prod}--~\eqref{eq:model_Pi}.

\section{Normalisation to extract the non-equilibrium spectrum and kinetic energy profile \label{app:Normalisation}}

In this Appendix, we investigate the different normalisation procedures mentioned in \S~\ref{subsec:Equilibrium}.
We state that the energy spectrum is decomposed into equilibrium (labelled by the subscript \(0\)) and nonequilibrium (labelled by the subscript \(1\)) contributions  as
\begin{equation}
  E(k, z) = E_0(k, z) + E_1(k, z).
\end{equation}
We further assume that the nonequilibrium contributions are zero-mean,
\begin{equation}
  \expval{ E_1(k, z) }_z = 0.
\end{equation}
We have therefore
\begin{equation}
  \expval{ E(k, z) }_z = \expval{ E_0(k, z) }_z.
\end{equation}
In order to compute the nonequilibrium contributions \(E_1(k, z) = E(k, z) - E_0(k, z)\),  we need to know the $z$-dependence of $E_0(k, z)$.
For this purpose, we use self-similarity assumptions and Kolmogorov's equilibrium hypothesis.


Scaling ranges in turbulence spectra appear when scale separation is attained, i.e., at sufficiently high Reynolds numbers.
In general, one can write the energy spectrum to scale as (see~\eqref{eq:E0_normalisation}),
\begin{equation}\label{eq:k41ff}
  E_0(k, z) \sim \epsilon(z)^{2/3} k^{-5/3} f_L[kL] f_\eta[k\eta(z)],
\end{equation}
at high-Reynolds numbers.
We have two nondimensional functions in~\eqref{eq:k41ff}; the \(f_L\) determines the shape of the spectrum for small \(k\) (large-scale) and \(f_\eta\) for large \(k\) (small-scale), respectively.
These functions satisfy the framework of~\citet{Kolmogorov1941_the_local},
\begin{equation}
  \lim_{x\rightarrow 0} f_\eta[x] = \lim_{x\rightarrow \infty} f_L[x] = 1.
\end{equation}
Therefore, we retrieve~\eqref{eq:E} for scales $L^{-1} \ll k \ll \eta^{-1}$ in the limit of $L/\eta \rightarrow \infty$.
Multiplying both sides of~\eqref{eq:k41ff} by $\eta^{-5/3}$ and dividing by $\epsilon^{2/3}$, we obtain
\begin{equation}\label{eq:E_Fketa}
  \frac{E_0(k, z)}{\epsilon(z)^{1/4} \nu^{5/4}} = F_\eta[k\eta(z)] f_L[kL],
\end{equation}
with
\begin{equation}\label{eq:F_keta}
  F_\eta[k\eta(z)] = (k\eta)^{-5/3} f_\eta[k\eta(z)].
\end{equation}
Since $F(kL)$ tends to unity for $k \gg 1/L$, the equilibrium spectra $E_0(k, z)$ should collapse when normalised by~\eqref{eq:E_Fketa} for any $z$, for large \(kL\).

Similarly, if the large scales are characterised by a length scale $L$, we can propose an alternative normalisation for~\eqref{eq:E_Fketa},
\begin{equation}\label{eq:E0_fkL}
  \frac{E_0(k, z)}{\epsilon(z)^{2/3} L^{5/3}} \sim F_L[k L] f[k\eta],
\end{equation}
with
\begin{equation}\label{eq:F_kL}
  F_L[k L] = (kL)^{-5/3} f_L[k L].
\end{equation}
It should scale purely as a function of $kL$ for $k\eta\ll 1$.

\begin{figure}
  \centering
  \includegraphics[width=\textwidth]{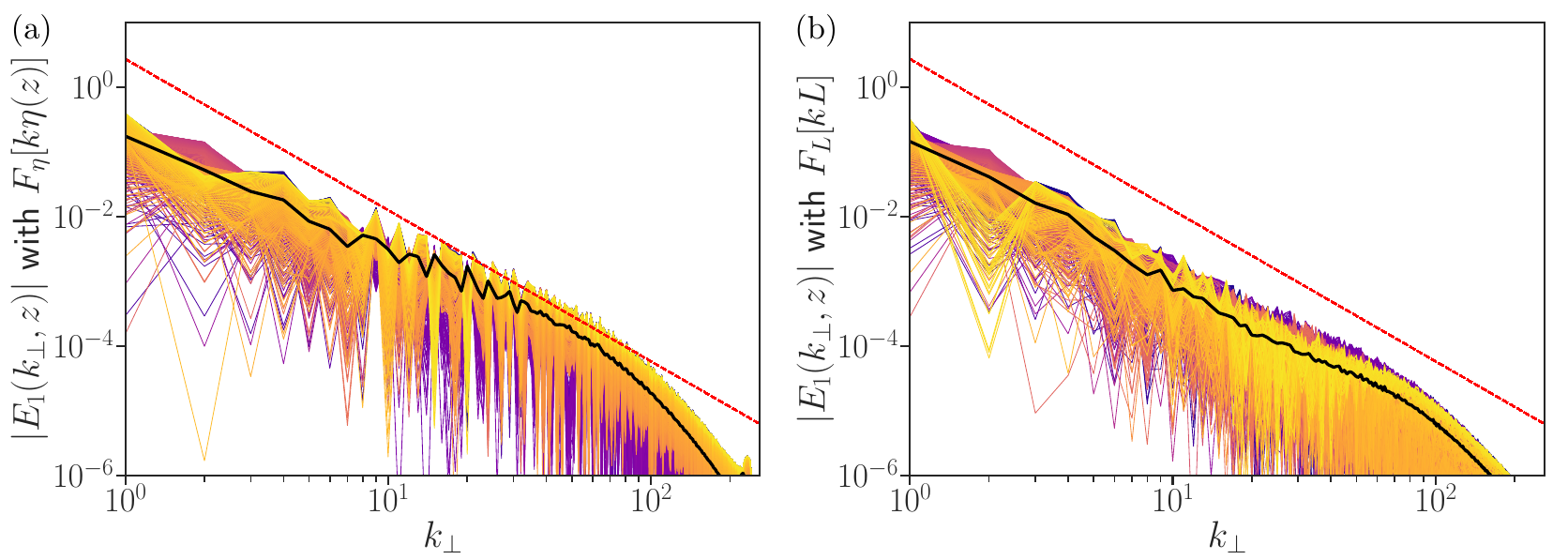}
  \caption{
    Absolute value of the time-averaged nonequilibrium energy spectrum \(\abs{E_1(k_\perp, z)} = \abs{\expval{E_1(k_\perp, z, t)}_t}\) for the highest Reynolds number dataset.
    Different nondimensionalised functions are employed to compute the nonequilibrium spectrum; panel~(a) with~\eqref{eq:F_keta} and~(b) with~\eqref{eq:F_kL}, respectively.
    The black solid and red dashed lines denote the \(z\)-average and \(k_\perp^{-7/3}\) scaling, respectively.
  }
  \label{Fig:E1_normalisation_dependency}
\end{figure}
There are, therefore, two normalisation possibilities.
One focuses on the high wavenumber limit of the inertial range close to the dissipation range~\eqref{eq:E_Fketa}--\eqref{eq:F_keta}, the other one on the low $k$ range close to the energy-range~\eqref{eq:E0_fkL}--\eqref{eq:F_kL}.
In the limit of infinite Reynolds number, we should find them to be equivalent in the inertial range, since
\begin{equation}
  \lim_{x\to\infty} F_L[x] = \lim_{x\to 0} F_\eta[x] = x^{-5/3}.
\end{equation}
Figure~\ref{Fig:E1_normalisation_dependency} plots the absolute value of the nonequilibrium energy spectrum \(E_1(k_\perp, z)\) for these two normalisations.
It is observed that using \(F[kL]\) we reveal a larger inertial range.
We will therefore use this normalisation in the present investigation.

Since the kinetic energy is dominantly determined by large scales, \eqref{eq:E0_fkL} allows us to determine the equilibrium kinetic energy profile,
\begin{equation}
  K_0(z)
    = \int E_0(k, z) \dd{k}
    = C_L \epsilon(z)^{2/3} L^{2/3}
  \label{eq:K0_eps23}
\end{equation}
with \(C_L = \int F_L [x] \dd{x}\).
Then, we define the decomposition
\begin{equation}
  K_0(z) = \expval{K(z)}_z + \widetilde{K}_0(z),
  \label{eq:K0_decomposition}
\end{equation}
where \(\expval{K_0 (z)}_z = \expval{K(z)}_z\) follows from the assumption that \(\expval{K_1(z)}_z = 0\).
By employing the decomposition for the energy dissipation rate profile
\begin{equation}
  \epsilon(z)^{2/3} = \expval{\epsilon(z)^{2/3}}_z + \widetilde{\epsilon(z)^{2/3}},
\end{equation}
it follows from~\eqref{eq:K0_eps23} that
\begin{equation}
  \frac{\widetilde{K}_0(z)}{\expval{K_0(z)}_z} = \frac{\widetilde{\epsilon(z)^{2/3}}}{\expval{\epsilon(z)^{2/3}}_z},
\end{equation}
and by~\eqref{eq:K0_decomposition},
\begin{equation}
  K_0(z) = \qty(1 + \frac{\widetilde{\epsilon(z)^{2/3}}}{\expval{\epsilon(z)^{2/3}}_z}) \expval{K(z)}_z.
\end{equation}
Since all the terms on the RHS are known, one can evaluate the nonequilibrium kinetic energy profile \(K_1(z) = K(z) - K_0(z)\) (see~\eqref{eq:K_decomposition}).

\backsection[Ackowledgements]{
  We dedicate this work to Jean-Pierre Bertoglio, on the occasion of his retirement.
  All DNS simulations were carried out using the facilities of the PMCS2I (\'Ecole Centrale de Lyon).
  R.A. is supported by the Takenaka Scholarship Foundation.\\
  For the purpose of Open Access, a CC-BY public copyright licence
  has been applied by the authors to the present document and will
  be applied to all subsequent versions up to the Author Accepted
  Manuscript arising from this submission.
}

\backsection[Declaration of interests]{The authors report no conflict of interest.}

\bibliographystyle{jfm}
\bibliography{biblio}
\end{document}